%% file: conference_101719.tex
\documentclass[conference]{IEEEtran}
\IEEEoverridecommandlockouts
\usepackage{cite}
\usepackage{amsmath,amssymb,amsfonts}
\usepackage{algorithmic}
\usepackage{graphicx}
\usepackage{textcomp}
\usepackage{amsmath}
\usepackage{xcolor}
\usepackage{tabularx}
\usepackage{pgfplots}
\usepackage{url}
\usepackage{hyperref}

\newcommand\blfootnote[1]{%
 \begingroup
 \renewcommand\thefootnote{}\footnote{#1}%
 \addtocounter{footnote}{-1}%
  \endgroup
}
 \makeatletter
\def\footnoterule{\kern-3\p@
  \hrule \@width 2in \kern 2.6\p@} 
\makeatother

\def\BibTeX{{\rm B\kern-.05em{\sc i\kern-.025em b}\kern-.08em
    T\kern-.1667em\lower.7ex\hbox{E}\kern-.125emX}}
\begin{document}

\title{SPIRE-SIES: A Spontaneous Indian English Speech Corpus}

\author{
Abhayjeet Singh\textsuperscript{1},
Charu Shah\textsuperscript{2}$^*$,
Rajashri Varadaraj\textsuperscript{2}$^*$,
Sonakshi Chauhan\textsuperscript{3}$^*$,
Prasanta Kumar Ghosh\textsuperscript{1} \\
\\
\textsuperscript{1} Department of Electrical Engineering, Indian Institute of Science (IISc), Bangalore-560012, India \\
\textsuperscript{2} National Institute of Technology Karnataka (NITK), Surathkal - 575025, India \\
\textsuperscript{3} United College Of Engineering And Research (UCER), Prayagraj - 211008, India\\
}

\maketitle
\blfootnote{$^*$These authors contributed equally to this work}
\begin{abstract}
In this paper, we present a 170.83 hour Indian English spontaneous speech dataset. Lack of Indian English speech data is one of the major hindrances in developing robust speech systems which are adapted to the Indian speech style. Moreover this scarcity is even more for spontaneous speech. This corpus is crowd-sourced over varied Indian nativities, genders and age groups. Traditional spontaneous speech collection strategies involve capturing of speech during interviewing or conversations. In this study, we use images as stimuli to induce spontaneity in speech. Transcripts for 23 hours is generated and validated which can serve as a spontaneous speech ASR benchmark. Quality of the corpus is validated with voice activity detection based segmentation, gender verification and image semantic correlation. Which determines a relationship between image stimulus and recorded speech using caption keywords derived from Image-to-Text model and high occurring words derived from whisper ASR's generated transcripts.

\end{abstract}


\section{Introduction}
\input{introduction}
\vspace{-0.2cm}
\section{Corpus Creation}
\subsection{Data Collection}
\input{audio_collection}

\subsection{Data Pre-processing}
\input{sorted_dataset}

\section{Dataset Visualisation}
\input{dataset_visualisation}

\section{Semantic Correlation of Utterances with Images}

\input{ASR}



\section{Potential Applications}
\subsection{Indian Nativity Classification}
\input{nativity_classification}

\subsection{Image to Indian English Speech Conversion}

\input{imagetoindianenglish}

\subsection{Spontaneous Indian English ASR}
\input{indian_nativity_class}

\subsection{ASR Benchmarking}
\input{ASR_Benchmarking}

\section{Conclusion}
\input{conclusion}


\bibliographystyle{unsrt}
\bibliography{references}

\end{document}

%% file: introduction.tex
A number of native English speech corpora are available for speech research and development. These corpora are useful for a number of applications including the development of speech technologies like Automatic Speech Recognition (ASR) and Text to Speech (TTS) systems. Some of the more popular read/prepared English speech corpora are LibriSpeech\cite{panayotov2015librispeech} (960 hours), TIMIT \cite{zue1990speech} ($\sim$5.5 hours), TED-LIUM 3\cite{hernandez2018ted} (452 hours), Wall Street Journal speech corpus (WSJ0 and WSJ1)\cite{paul1992design} (80 hours) etc. On the other hand in the category of Spontaneous English, there are Switchboard corpus\cite{godfrey1992switchboard} (250 hours), Fisher Corpus\cite{cieri2004fisher} (2000 hours), Buckeye\cite{pitt2005buckeye} ($\sim$40 hours), LJ speech (24h) \cite{ljspeech17}, CMU-Artic ($\sim$1+ hour)\cite{article2}, The M-AILABS Speech (multi-speaker) (292 hours) \cite{bakhturina2021hi}, Blizzard Challenge 2011 (16.6 hours) \cite{King2011TheBC}, The World English Bible \cite{resnik1999bible}, LibriTTS (multi-speaker) (585 hours)\cite{zen2019libritts} etc. 

Even with the abundance of read and spontaneous English speech resources, the performance of speech-based tasks doesn't translate well when applied to Indian English speech \cite{9660500}. This is due to influence of first language (L1) on second (L2) and third (L3) language \cite{hammarberg2001roles, sinha2009interference}. Thus, to achieve good performance on Indian English speech tasks, data from non-native English recordings from native Indian speakers is necessary. In the past, there have been many advancements towards open-sourcing Indian English speech corpora, like Indic TIMIT \cite{9041230} ($\sim$240 hours) in read speech category, Svarah \cite{javed2023svarah} (9.6 hours) having both read as well as spontaneous speech and NPTEL2020 ($\sim$15.7K hours) \cite{nptel2020} which could fall under prepared speech category. These statistics indicate the unavailability of a good amount of Indian English data in spontaneous speech category, for speech research and development.

 Typically in a real-world application several speech-based systems, e.g.,  ASR, are expected to receive spontaneous speech. Hence, it is imperative that they perform well in spontaneous speech. However, it has been noted that the performance of the ASR system deteriorates when the style of speech changes from read to spontaneous. This degradation occurs when the ASR system, originally trained on scripted speech, is utilized for transcribing spontaneous speech and vice-versa. Notably, the ASR performance exhibits enhancement when both the training and testing data correspond to the same speech type.\cite{butzberger1992spontaneous}. This suggests that the development of an English ASR in the Indian context would demand Indian English spontaneous speech data.

Data-collection strategies for read speech are straightforward i.e., the speaker reads and records a predetermined set of transcripts. But, in the case of spontaneous speech various protocols have been undertaken in the past for capturing spontaneity in speech. Interviewing the speakers on everyday topics such as politics, sports, traffic, schools \cite{pitt2005buckeye} and recording of telephonic conversations where questions or diverse topics are given as stimuli to the speakers\cite{cieri2004fisher, godfrey1992switchboard}. Unlike these, for the spontaneous speech collection in this work, we have used images from various domains as stimuli to the speakers rather than an array of topics/questions to speak on. The information content in an isolated topic or a single question is limited, thus the speech content from the speaker is dependent on the imagination and experience of the speaker. On the other hand, images are information rich and contextually bounded which guides the speaker and makes it easy for a speaker to speak. This results in a spontaneous, contextually rich, and semantically correlated spoken corpus. 

English being an L2 language in the Indian community, effects of the L1 language tend to have a major impact on how English is spoken by different nativities \cite{hammarberg2001roles, sinha2009interference}. With our image-stimulated spontaneous speech collection strategy, where multiple speakers from varied nativities speak about a common set of images, variation of linguistic content within semantically correlated utterances from different Indian nativities is captured. This could be potentially used to improve the performance of Indian English ASR and Spontaneous TTS systems.

Taking a step to meet the demand for Spontaneous Indian English Speech corpus, we are open-sourcing $170.83$ hours of spontaneous speech data in the English language corpus, called SPIRE-SIES spoken by Indian native speakers. In this corpus, we cover 12 major Indian nativities. Such a spontaneous speech corpus from multiple nativities will also aid the development of the Indian nativity classifier and detection systems. With data collected from 1607 speakers, pre-trained speaker identification models can be fine-tuned with this corpus to adapt to Indian English. With its unique image-based recording process, utterances for an image are semantically correlated and can be used to train Image to Indian English speech conversion models. Among these $170.83$ hours of speech, transcripts for 23 hours are also open-sourced. This 23 hours of transcribed Indian English spontaneous speech can prove to be a bench-marking test corpus for Indian English ASR systems. We are also working towards providing manually verified transcripts for the entire $170.83$ hours. Using this, the performance of the Indian English ASR system could be improved. 

%% file: audio_collection.tex
\begin{figure}[!h]
    \centering
    \includegraphics[width=0.75\linewidth]{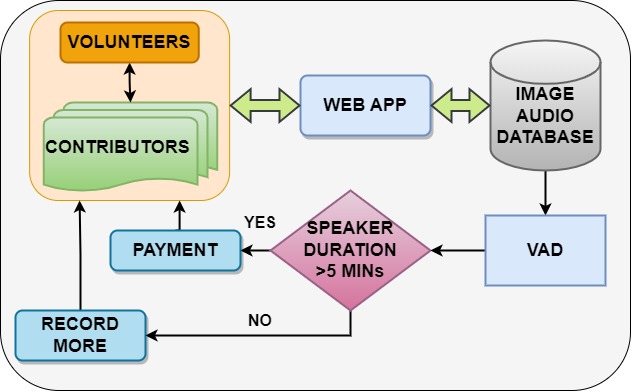}
    \caption{A flow diagram of data-collection process}
    \label{fig: - Flowchart from web_app to payment(including volunteers and contributors)}
    \vspace{-0.5cm}
\end{figure}

In this paper we present, SPIRE-SIES, a unique Indian English spontaneous speech corpus, which was crowd-sourced using a web application \footnote{https://native-audio-recorder.web.app} developed by SPIRE Lab, leveraging the React Native framework \cite{React}.
Various steps of the data collection process are shown in Fig \ref{fig: - Flowchart from web_app to payment(including volunteers and contributors)}. To begin with, Google forms were circulated in several Indian colleges/universities looking for student volunteers. The volunteers were instructed to find out 45 contributors and collect at least 5 mins from each of the 45 contributors, who were fluent in English. The contributors had to use the web app for this process and speak about the image shown on the screen. To record in the web app, contributors could log in using their Google/Facebook accounts or contribute anonymously. When recording for the first time in the web app, contributors were asked to provide essential metadata like name, year and month of birth, year of study, origin (from a drop-down list of states), native language/mother tongue (from a drop-down list of 22 Indian languages) and gender. Contributors were compensated only after a minimum of 5 minutes of effective voiced content was detected by a voice activity detection (VAD) algorithm \footnote{https://github.com/eesungkim/Voice\textunderscore Activity\textunderscore Detector} \cite{sohn1999statistical}\cite{1164453} from each of the 45 contributors. Even the volunteers were paid to ensure reliable recording as it was incentive-based.

\begin{figure}[h!]
\centering
  \includegraphics[width=0.8\linewidth]{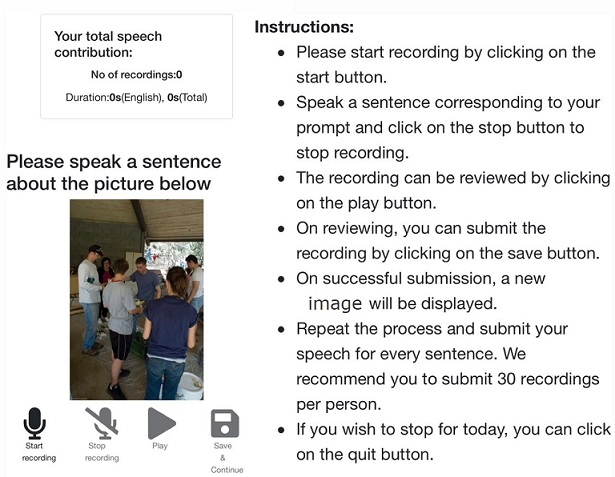}
  \caption{Screenshot of recording page with instructions from web app used for data collection}
  \label{fig: - Screenshot of the Web Application}
\end{figure}

\begin{figure}[h!]
\centering
  \includegraphics[width=0.3\textwidth]{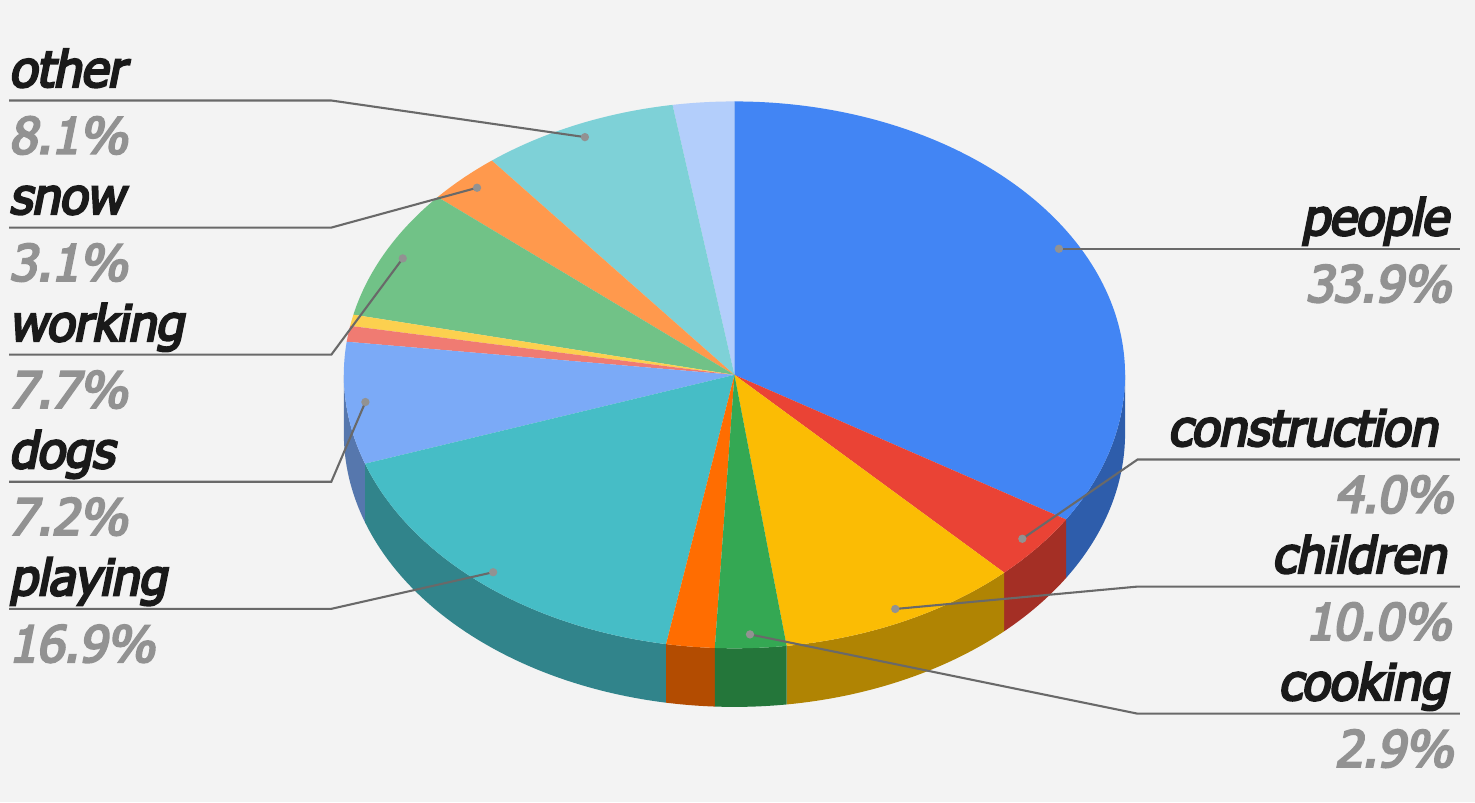}
  \caption{Domain distribution of images used as stimuli in SPIRE-SIES}
  \label{fig: - Classgenre}
\end{figure}

Fig \ref{fig: - Screenshot of the Web Application} shows a sample recording screen on the left along with an instruction set shown to the contributor. The recording screen in the web app features an image-based audio collection interface as shown in Fig \ref{fig: - Screenshot of the Web Application}. The recording screen consists of 3 components: 1) At the top, a recording statistics box displays the number and total duration of recordings by the contributor. 2) In the middle an image is displayed with a recording console with ``Start recording", and ``Stop recording" buttons to turn the microphone on and off, respectively, The ``Play" button plays the last recording by the contributor before saving by clicking the ``Save \& Continue" button. 3) At the bottom we have a set of instructions for the contributors as shown on the right side of Fig \ref{fig: - Screenshot of the Web Application}. Contributors need not record for 5 minutes in a single session as they could logout at any point and re-login to resume their work.  During the recording process, images were presented one by one from a pool of 1000 images, and a contributor was asked to speak about the image. 
Spontaneous speech was elicited by instructing users to describe images rather than reading prepared texts. A contributor could skip an image if he/she were not comfortable describing it. These 1000 images were selected from Flickr30k\cite{young2014image} image captioning dataset to cover various domains, resulting in a diverse set of recordings. Only 1000 images were selected as stimuli so that decent number recordings of each image could be collected. Fig. \ref{fig: - Classgenre} represents a  categorical domain distribution of the image stimuli set. 

The initial data collection phase encompassed a diverse audience from different Indian states and languages, yielding a substantial dataset of duration 170+ hours comprising 37895 samples. The focus was on capturing speech from speakers with 12 major Indian languages, reflecting various linguistic backgrounds and nativities. After pre-processing and VAD checks, 37505 samples from 1607 speakers were found to be suitable for the corpus. The data was obtained from different regions across India, ensuring geographical diversity and representation of native speakers. Each recording was uniquely identified by assigned speaker IDs, facilitating data organization and tracking. 

\vspace{-10pt}



%% file: sorted_dataset.tex
As the data collection is crowd-sourced and recorded in an uncontrolled environment, multiple pre-processing steps were implemented to enhance the quality of the dataset. This section details the pre-processing steps used in the SPIRE-SIES corpus having a total of 37,895 samples collected over the span of three years from 2021 and 2023. Although the data collection process had $1000$ images, recordings from 761 images were obtained. This could be because contributors skipped several images that were difficult to be described by them.


 The dataset was accompanied by a JSON metadata file containing essential speaker information such as ID, gender, and nativity, which were utilized to structure the data for subsequent analysis. To establish consistency, recordings from iOS and Android devices with different encodings were modified to adopt a common format. During the initial data assessment, 39 redundant samples were identified due to missing encodings or corrupted files. The raw recorded audio files were formatted to a single channel, 16-bit, and 16K sampling rate wav files.

To ensure data quality, the VAD algorithm was applied to detect speech and silence segments(every 10ms) in the recordings. The silence-to-voice ratio was calculated for each recording, enabling the identification of files with substantial silence content. Files with silence percentages of 75\% and above underwent manual review, resulting in the removal of $351$ samples. This curation process yielded a dataset comprising valid and pertinent audio samples, providing a substantial $170.83$ hours of audio content.

\begin{figure}[h!]
  \includegraphics[width=\linewidth, height= 4.5cm]
  {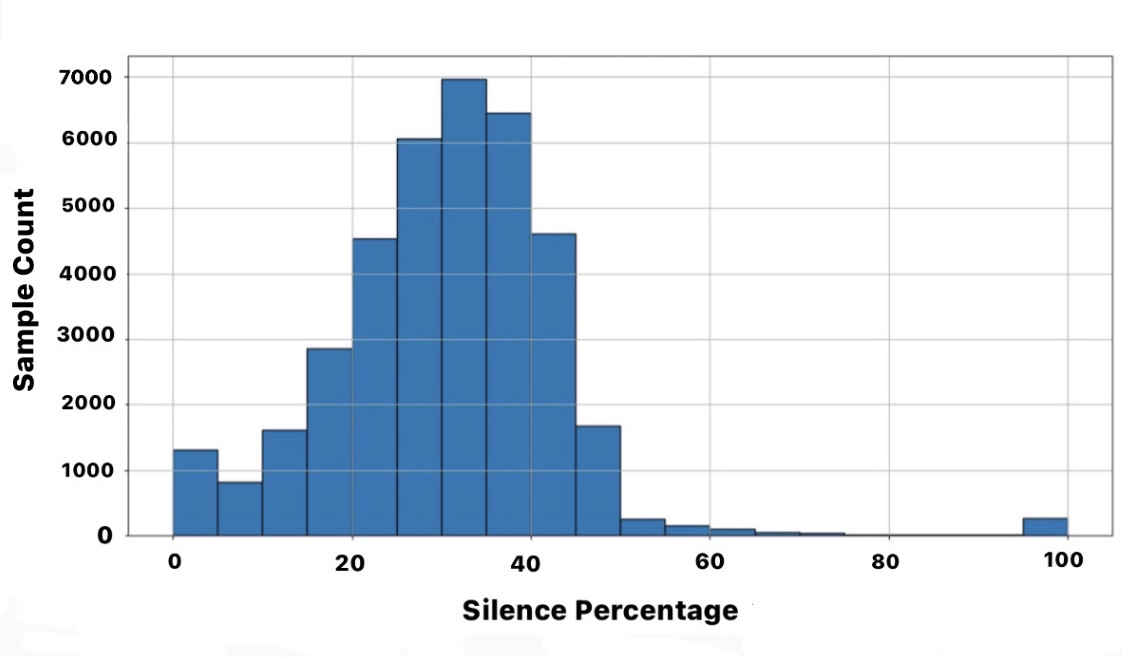}
  \caption{Frequency of files per percentage silence in a file}
  \label{fig:frequency of files versus percentage silence ratios.}
  \vspace{-0.5cm}
\end{figure}

The distribution of silence across the dataset is visually represented in Fig \ref{fig:frequency of files versus percentage silence ratios.}, displaying the histogram of silence percentages in recordings, which provides insights into the silence distribution in SPIRE-SIES.   

\begin{figure}[h!]
\centering
  \includegraphics[width=0.8\linewidth,trim=19mm 5mm 3mm 4mm,clip]{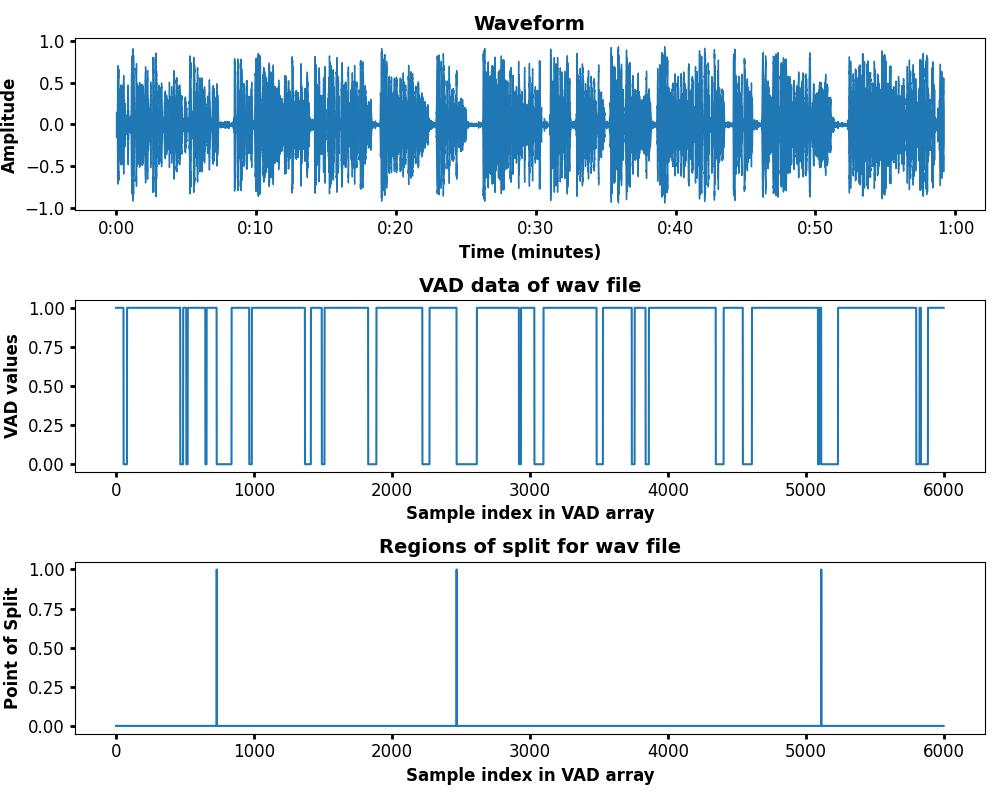}
  \caption{Segmentation performed on a sample recording}
  \label{fig:Segmentation performed on a sample recoriding}
  \vspace{-0.5cm}
\end{figure}

A total of 1203 files, the duration of each exceeding 60 seconds, undergo further VAD analysis. These files are segmented at points with nearly one second of silence before and after the split point. If a point of split is present within 10 seconds of another point, then the segmentation is done only at the first point. The segmentation performed on a single audio file is illustrated in Fig \ref{fig:Segmentation performed on a sample recoriding} as an example. The figure depicts the waveform, followed by a plot of VAD values obtained for the same. The presence of voice activity indicated by value $1$ and silences by value $0$ are plotted as the VAD values on the y-axis. The last subplot depicts the points of split for the illustrated waveform where the sample indices at which the file is to be segmented are marked by value $1$. 
The visual representation provided in Fig \ref{fig:Sample per duration range before and after segmentation} illustrates the possible impact of data segmentation on optimizing the dataset. This optimization strategy would lead to a considerable increase in the number of samples, as evidenced by the elongated \textit{blue} bars assigned to each speaker. The bars, symbolizing the sample count, serve as a clear testament to the successful handling of excessively long durations. This approach not only enhances the dataset's overall quality but also ensures a more balanced and representative distribution of samples across various duration ranges.

Additional filtering resulted in a more refined dataset that was obtained after the removal of 1) files that were unreadable, 2) files that had encoding problems that were primarily caused by the language being different from English or by the presence of two voices in the audio, one of which was the primary voice and the other distant and not in English, 3) files with transcription having encoding problems in UTF-8, and lastly, 4) files on which automatic transcription could not be done. This results in a reduced duration of 162.60 hour from the the original 170.83 hour.

%% file: dataset_visualisation.tex
To explore the correlations between different aspects of the data and its associated metadata, as well as gain valuable insights into the dataset's overall composition and diversity, several experiments were conducted. These studies and their findings are elaborated upon in the subsequent sub-sections.

\subsection{Speaker and Nativity analysis}
The dataset consists of $37,505$ samples from $1,607$ speakers. Since each speaker is associated with multiple samples, the statistics for gender distribution of both samples and speakers are presented in Table \ref{table:1}. The table depicts that the data is gender balanced. Others are the users who did not specify their gender. The number of images described and the number of samples for every language are indicated in Table \ref{table:2}. It can be inferred from these statistics that while Marathi has the maximum number of samples, the total time duration is highest for Telugu. Thus, Telugu speakers spoke for more duration than Marathi speakers though they were fewer in number.

\begin{table}[h!]
\centering
\begin{tabular}{ | m{3cm} | m{1cm}| m{1cm} | m{1cm} |} 
  \hline
    & Female
 & Male & Other\\ 
  \hline
  Percentage of samples & 45.375 & 54.230 & 0.395\\ 
  \hline
  Percentage of speakers & 48.628 & 50.873 & 0.499\\ 
  \hline
\end{tabular}
\vspace{2mm}
\caption{Table depicting gender distribution in SPIRE-SIES corpus.}
\label{table:1}
\vspace{-0.5cm}
\end{table}

 \subsection{Total Time Duration and Duration Range Analysis}  The total time duration spoken in minutes per language was computed (Table \ref{table:2}). The average time duration spoken per sample for each nativity shows that samples belonging to Telugu nativity have the longest average duration of about $0.5374$ minutes. This is more than double the average duration computed considering all nativities, which is around $0.2447$ minutes. Konkani has the least average duration per sample, of $0.1192$ minutes. This analysis provides an understanding of the distribution of speaking time across different languages, aiding in identifying any variations or patterns in the data1..

 \begin{table}[h!]
\centering
\resizebox{\linewidth}{!}{
\begin{tabular}{ | m{1.5cm} | m{1cm}| m{1cm} | m{1cm} | m{1.5cm} | m{1cm} |} 
  \hline
   Nativity & Number of Samples
 & Number of Images & Duration in minutes& Average duration (minutes) per sample & Average age (years) per nativity\\ 
  \hline
  Telugu & 5883 & 748 & 3163.805 & 0.5378 & 20\\ 
   \hline
  Hindi & 6687 & 626 & 1797.538 & 0.2688 & 21\\ 
   \hline
  Marathi& 7011 & 666 & 1600.111 & 0.2282 & 21\\
  \hline
  Tamil & 3604 & 619 & 836.223 & 0.2320 & 21\\ 
   \hline
  Malayalam & 3070 & 650 & 792.337 & 0.2581 & 22\\ 
  \hline
  Kannada & 3641 & 642 & 608.779 & 0.1672 & 21\\ 
  \hline
  Bengali & 3522 & 694 & 605.115 & 0.1718 & 22\\   
   \hline
  Odia & 803 & 520 & 193.021 & 0.2404 & 22\\ 
  \hline
  Konkani & 1122 & 574 & 133.754 & 0.1192 & 23\\ 
  \hline
  Gujarati & 146 & 135 & 49.715 & 0.3405 & 20\\ 
  
   \hline
  Punjabi & 221 & 187 & 40.509 & 0.1833 & 21\\ 
  \hline
  Maithili & 195 & 164 & 36.732 & 0.1884 & 24\\ 
  \hline
  Others & 1600 & 667 & 392.211 & 0.2451 & 22\\ 
  \hline
\end{tabular}
}
\vspace{2mm}
\caption{Table depicting nativity of recordings in SPIRE-SIES corpus.}
\label{table:2}
\end{table}

\begin{figure}[h!]
\centering
\includegraphics[width=0.75\linewidth,trim=10mm 4mm 10mm 5mm,clip]{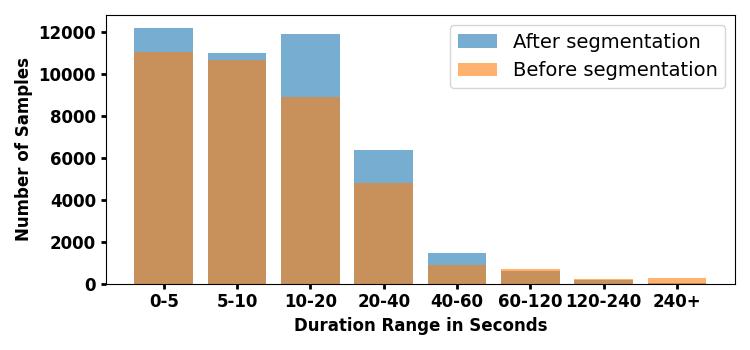}
  \caption{Histogram of duration range before and after segmentation}
  \label{fig:Sample per duration range before and after segmentation}
  \vspace{-0.5cm}
\end{figure}

To examine data quality and potentially remove outliers, we plotted the sample count per duration range for the entire dataset as shown in Fig. \ref{fig:Sample per duration range before and after segmentation}. This helped in identifying excessively long or short recordings, aiding in potential data cleaning and filtering processes. Segmentation performed using VAD as explained in the previous section, would yield differences in distribution as indicated in Fig. \ref{fig:Sample per duration range before and after segmentation} itself. The audio samples with a duration greater than 60 seconds, when segmented into smaller audio samples, at points of silence, lead to an increase in the number of samples for ranges within 60 seconds.

\subsection{Gender verification}
The audio data was crowd-sourced. Hence, the gender mentioned by the user had to be verified. For this purpose a pre-trained model \cite{Gender} was used to find the gender. The model was trained on a combined set of audios in Hindi, Tamil, Telugu, and Kannada languages and comprised of  $17.4$ hours of male and $16.2$ hours of female audio. For testing, the languages of Hindi, Tamil, Telugu, Kannada, Marathi and Bengali were used. The final test dataset had a distribution of $3.6$ hours for both males and females. The SPIRE-SIES corpus was then tested using this model and the confusion matrix mentioned in Table \ref{cf} was obtained. The output obtained on the corpus data had an F1 Score of $96.047\%$. 
\begin{table}[h!]
\centering
\begin{tabular}{|l|l|l|l|}
\hline
                     & \textbf{Male}         & \textbf{Female}  \\ \hline
\textbf{Male} & \multicolumn{1}{c|}{19385} & \multicolumn{1}{c|}{954}    \\ \hline
\textbf{Female}   &          524                &     16494                     \\ \hline
\end{tabular}
\vspace{2mm}
\caption{Confusion Matrix of Corpus Data}
\label{cf}
\vspace{-0.5cm}
\end{table} 
Based on the information derived from Table \ref{cf}, it becomes evident that among the total of $37,357$ cases, there were $1,478$ instances where the gender specified by the user did not align with the model's predictions. This discrepancy accounted for only a 3.96\% error rate. Upon closer examination of these cases, it was found that they were consistent with the gender information provided by the users. Consequently, these cases were considered valid and retained for further analysis.

%% file: ASR.tex
Before proceeding with the $36,014$ recordings, the relevance of the speech content in these recordings in context to the image stimuli had to be determined. Thus filtering out the audio files, which actually correspond to the image. To analyze and gain insights from the dataset, a pre-trained ASR is used to estimate text from audio samples. We utilized Whisper\cite{OpenAIWhisper} ASR by OpenAI for the speech-to-text conversion, as it is one of best performing ASR systems currently available for English language.

\begin{figure}[h!]
    \centering
    \includegraphics[width=7cm,trim=0mm 6mm 0mm 0mm,clip]{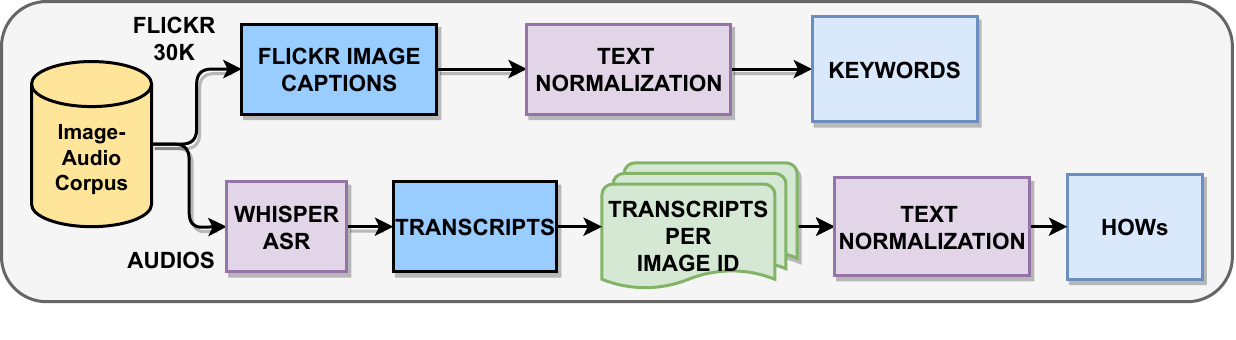}
    \caption{Extraction of HOWs and Keywords.}
    \label{fig:ex_key}
\end{figure}

After performing text pre-processing i.e., stop-word removal, symbols removal, and stemming on the Whisper transcripts, we extract a contextually rich text. This text was then utilized for determining a descriptive set of keywords for each image, namely \textit{High Occurring Words(HOWs)}. The process for which is shown in Fig. \ref{fig:ex_key}. All the transcripts pertaining to a specific image spoken by various speakers were aggregated and subjected to pre-processing. Subsequently, we identified the top 5 words that consistently appeared across all these transcripts, taking into account all speakers. These frequently occurring words, which we refer to as HOWs, represent the vocabulary shared by all audio samples associated with that particular image. This approach ensures that the selected words are those that appear most frequently across all audio samples, collectively from all speakers, providing a comprehensive understanding of the image's content.

The $1000$ image set which is used as stimuli in our corpus is taken from the Flickr30K dataset \cite{flickr30k}  which comprises $31K$ diverse images, each paired with $5$ reference captions. This dataset was chosen for its diversity in images and its multiple captions per image facilitated a nuanced exploration of visual-linguistic relationships. These $5$ set image captions are being utilized by us in our further analysis for which the process of keyword extraction is shown in Fig. \ref{fig:ex_key}, the $5$ set captions given for each image from the original database were taken and after text normalization unique words were taken as the keywords.
\vspace{-4mm}
\vspace{-4mm}

In order to determine the semantic relationship between the spontaneous recordings and the image stimuli, we quantified two degrees of overlap between: 1) Whisper transcripts and the Flickr captions keywords (Caption\_Overlap\_Percentage) 2) Whisper transcripts and the HOWs (HOWs\_Overlap\_Percentages). Below are formulae to calculate these overlap percentages:

\begin{multline}
\textbf{Caption\_Overlap\_Percentage} = \\
 \frac{Count\ of\ HOWs\ in\ Whisper\ transcript}{Keywords\ in\ the\ Whisper\ transcript} 
\end{multline}
\begin{multline}
\textbf{HOWs\_Overlap\_Percentage} = \\
 \frac{Count\ of\ Flickr\ Keywords\ in\ Whisper\ transcript}{Keywords\ in\ the\ Whisper\ transcript}
\end{multline}

We employed a simple logic pertaining to the fact that if the Keywords or the Hows are not present in the transcript, the image is not semantically related. 
Using the Caption and HOWs overlap percentages, we found that only 13.6\% (4989 recordings) were semantically unrelated to their corresponding image stimuli i.e., for these recordings both our overlap metrics were 0\%. And a correlation of 0.69 between the 2 overlap metrics further bolsters their validity and consistency.

Thus a total of $86.4\%$ of the Whisper transcripts exhibited a meaningful semantic alignment with the corresponding image captions. This finding underscores a compelling conclusion: Whisper transcripts maintain a significant level of relevance to the underlying images, thus validating their utility within our research context.

%% file: nativity_classification.tex
A Spontaneous Indian English Speech Corpus holds great significance in addressing the nativity classification problem within Indian English speech. Hence, SPIRE-SIES corpus is instrumental in developing machine learning models capable of recognizing diverse accents and dialects in Indian English, facilitating the accurate identification of a speaker's regional background based on pronunciation patterns. 

%% file: imagetoindianenglish.tex
There are already many pre-existing speech corpora \cite{10023089,elliott2016multi30k} having image and audio descriptions of the respective image serving as an image-to-speech dataset. 
The spontaneous collection in SPIRE-SIES corpus caters to Indian listeners, delivered in Indian English being recorded in Indian English. Most of the existing corpora are in English or other high-resource languages. This paper attempts to address the neglected part of how English and Indian English are different and contribute for the same.

%% file: indian_nativity_class.tex
The development and implementation of a specialized Spontaneous English Indian ASR system, as discussed in the papers by Ramabhadran et al. \cite{ramabhadran2009robust} and Sharma et al. \cite{sharma2020effect}, yields significant benefits across diverse sectors. This system, trained on a comprehensive dataset of spontaneous Indian English speech, effectively addresses the challenges posed by Indian accents and linguistic variability. Spontaneous English Indian ASR system's ability to capture natural variations in Indian English and its adaptability to real-world scenarios. SPIRE-SIES corpus enables development of such spontaneous Indian English ASR.


%% file: ASR_Benchmarking.tex

Several corpora \cite{ kolobov2021mediaspeech, mihajlik2022beabase, 9997917, fujimotoreazonspeech } have been specifically curated for the purpose of advancing ASR technology. These datasets are tailored to particular domains or are aligned with specific language groups. Our analysis has identified a noticeable void in available datasets, prompting us to meticulously compile and create this dataset with the intention of facilitating accurate and comprehensive ASR benchmarking. 23 hour subset from SPIRE-SIES corpus can be effectively used for ASR benchmarking. 



%% file: conclusion.tex
SPIRE-SIES corpus presented in this work is a 170 hour open-sourced spontaneous Indian English speech dataset. Out of which 23 hours of data is currently transcribed. This corpus includes data of subjects from multiple Indian nativities. In order to access SPIRE-SIES corpus, please submit a request through this \href{https://forms.office.com/r/bF27ah1Fg9}{\textcolor{blue}{form}}. While this corpus, for the first time, fills the unavailability of spontaneous Indian English speech corpus, the duration of SPIRE-SIES is not significantly large. Thus, future works entail further collection of spontaneous English speech from Indian speakers, particularly from different age groups, educational backgrounds. Along with exploratory studies on various modes of stimuli for spontaneity in speech.